\documentclass[conference]{IEEEtran}

\usepackage{amsmath}
\usepackage{amssymb}
\usepackage[dvips]{graphicx}
\usepackage{setspace}
\usepackage{epsfig}
\usepackage[compress]{cite}
\usepackage{subfigure}
\usepackage{algorithm,algpseudocode,caption}
\usepackage{color}
\usepackage{lipsum}

\newcommand{\tSNR}{\text{SNR}}

\newtheorem{prop}{Proposition}

\newcommand{\figsizee}{0.31}

\begin{document}
\IEEEoverridecommandlockouts
\title{Performance Analysis of Energy-Detection-Based Massive SIMO}
\author{\authorblockN{Marwan Hammouda, Sami Akin, and J\"{u}rgen Peissig}
\authorblockA{Institute of Communications Technology\\
Leibniz Universit\"{a}t Hannover\\
Email: \{marwan.hammouda, sami.akin, and peissig\}@ikt.uni-hannover.de}}	
\date{}

\maketitle
\pagestyle{empty}

\begin{abstract}
Recently, communications systems that are both energy efficient and reliable are under investigation. In this paper, we concentrate on an energy-detection-based transmission scheme where a communication scenario between a transmitter with one antenna and a receiver with significantly many antennas is considered. We assume that the receiver initially calculates the average energy across all antennas, and then decodes the transmitted data by exploiting the average energy level. Then, we calculate the average symbol error probability by means of a maximum a-posteriori probability detector at the receiver. Following that, we provide the optimal decision regions. Furthermore, we develop an iterative algorithm that reaches the optimal constellation diagram under a given average transmit power constraint. Through numerical analysis, we explore the system performance.
\end{abstract}

\section{Introduction}
The steadily growing wireless data traffic is a key reason behind the research for new transmission technologies with high performance gains and high energy efficiency. Among these technologies, massive multiple-input multiple-output (MIMO) systems, also called Large-Scale Antenna systems \cite{Special_Issues}, made a dramatic strike since they are energy efficient and reliable \cite{Liu}. The concept of Massive MIMO, initially proposed in \cite{Marzetta2010}, is based on using significantly many transmit and/or receive antennas, which leads to remarkable advances in spectral efficiency, beamforming gain and radiated energy efficiency \cite{Special_Issues}, \cite{ngo2013energy}. However, the large number of antennas poses a major challenge for massive MIMO to become a reality, since obtaining channel side information (CSI) gets more ponderous \cite{hoydis2011massive}. Therefore, massive MIMO is applicable within time-division-duplex (TDD) systems, since channel reciprocity can be utilized \cite{hoydis2011massive}. However, when applied in multi-cell systems such as cellular networks, massive MIMO within TDD systems leads to another problem, which is called pilot contamination. This is due to that the number of orthogonal pilot tones is limited in each cell, and these orthogonal pilot tones are reused across cells \cite{Jose2009}.   

To overcome the problem of pilot contamination, a considerable research effort has been expended, and several solutions have been proposed. One of these solutions is to design systems that do not require CSI at either the receiver or the transmitter. In this context, a fundamental work is presented in \cite{Unitary,Grassman,Sugumar2007,Shamai} where space-time coding over the Grassman manifold associated with the channel matrix is performed. However, these studies are based on the assumption of high signal-to-noise-ratio availability. In a more recent study \cite{chowdhury2014design}, the authors considered a noncoherent single-input multiple-output (SIMO) system with a large number of receive antennas, and proposed a simple energy-detection-based encoding and decoding scheme. They derived an upper bound for the average symbol error probability of the proposed scheme, and they showed through simulations that this upper bound will vanish exponentially with the increasing number of receive antennas. The authors also proposed a simple constellation design based on the minimum distance criterion. 

In another line of research, the general problem of maximizing the system performance by deriving optimal design schemes is researched in numerous studies \cite{wang2007approximate, goldfeld2002minimum,Ding2011}. For example in \cite{wang2007approximate}, the authors proposed a power allocation strategy that minimizes the bit error rate (BER) in MIMO spatial multiplexing systems. On the other hand, the authors in \cite{goldfeld2002minimum} developed an optimal power allocation strategy for orthogonal frequency-division multiplexing systems in order to minimize the cumulative BER. In all of these studies, it is assumed that both the transmitter and the receiver have perfect knowledge of CSI. 

In this paper, we consider an energy-detection-based communications system in which a transmitter with one antenna communicates with a receiver with many antennas. We investigate a modulation and demodulation technique based on the calculation of energy across all receive antennas. Obtaining the average symbol error probability after a maximum a-posteriori probability (MAP) detector is employed, we provide the optimal decision regions. We develop an iterative algorithm that converges to the optimal constellation diagram under a given average transmit power constraint.\footnote{In this paper, we identify the optimal constellation diagram that minimizes the exact average symbol error probability rather than an upper bound to it as considered in \cite{chowdhury2014design}.}

\section{System Model}
\begin{figure}
    \centering
    \subfigure[Channel scenario.]
    {\includegraphics[width=0.36\textwidth]{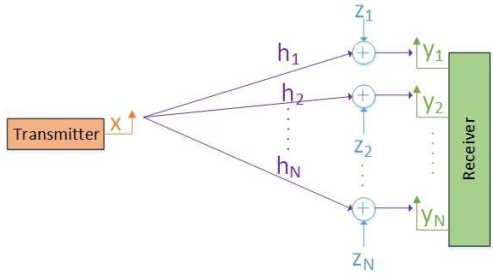}\label{System_Model}}
		\\
    \subfigure[Constellation diagram.]
    {\includegraphics[width=0.36\textwidth]{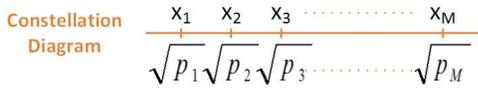}\label{Constellation}}
    \label{fig:sample_subfigures}
		\caption{System model.}
\end{figure}

As shown in Figure \ref{System_Model}, we consider a communications scenario in which one transmitter having a single antenna performs data transmission to one receiver that has $N$ antennas. During data transmission, the input-output relation is given by
\begin{equation*}
{\bf y} = {\bf h} x + \boldsymbol{z},
\end{equation*}
where $x$ is the data symbol sent by the transmitter and ${\bf y}$ is the $N\times1$-dimensional output vector at the receiver. Above, $\boldsymbol{z}$ is the $N\times1$-dimensional additive noise vector. Each of its elements, $z_{n}$, is a zero-mean Gaussian random variable with variance $\sigma_{z}^{2}$ for $n\in\{1,\cdots,N\}$. Meanwhile, $\boldsymbol{h}$ represents the $N\times1$-dimensional channel vector, each element of which, $h_{n}$, is also a Gaussian-distributed random variable but with mean $\mu$ and variance $\sigma_{h}^{2}$. We also consider the following dynamics for $h_{n}$: $|\mu|^2 = \frac{K}{K+1}$ and $\sigma_h^2 = \frac{1}{K+1}$ for a known real number $K$ where $0\leq K$. Note that this is the Rician fading channel model \cite{goldsmith2005wireless}. We further assume that neither the transmitter nor the receiver knows the instantaneous realizations of the channel. However, they are aware of the system statistics such as $\sigma_z^2$ and $K$. 

Since an energy-detection-based encoding and decoding technique is considered, the transmitter sends real positive symbols from a constellation of $\mathcal{P}=\{\sqrt{p_1},\cdots,\sqrt{p_{M}}\}$ as shown in Fig. \ref{Constellation} rather than transmitting complex symbols. $M$ is the constellation size and $p_m$ is the power level of the $m^{th}$ symbol, $x_{m}$, i.e., $x_{m}=\sqrt{p_{m}}$, for $m\in\{1,\cdots,M\}$. We note that $0\leq p_{1}<p_{2}<\cdots<p_{M}<\infty$. Moreover, assuming that each symbol is sent with equal probability $\frac{1}{M}$, we impose the following average power constraint: 
\begin{equation}\label{int_constraint}
	\frac{1}{M}\sum_{m=1}^{M}p_m \leq \bar{p},
\end{equation}
where $\bar{p}$ is the average symbol power. Then, we define the signal-to-noise ratio as $\tSNR = \frac{\bar{p}}{\sigma_z^2}$. Finally, we denote the ratio of the power of the $m^{th}$ transmitted symbol to the average power by $\alpha_m = \frac{p_m}{\bar{p}}$.
\section{Average Symbol Error Probability}
\begin{figure}
\begin{center}
\includegraphics[width=0.49\textwidth]{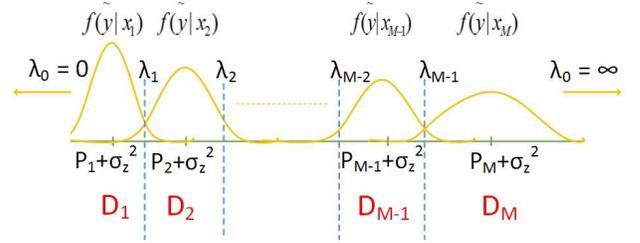}
\caption{Decision regions and boundaries for decoding.}\label{Received_Constellation}
\end{center}
\end{figure}

We assume that in order to decode the transmitted symbol, the receiver utilizes the average received energy across all antennas, i.e., $\widetilde{y}=\frac{\|{\bf y}\|^2}{N}.$ We can express the average received energy across all antennas when $x_{m}$ is transmitted as
\begingroup
\allowdisplaybreaks
\begin{align}
	\widetilde{y}&=\underbrace{\frac{\|{\bf h}\|^2}{N}p_m}_{p_{m}+w_{m,1}}+\underbrace{\frac{\|\boldsymbol{z}\|^2}{N}}_{\sigma_{z}^{2}+w_{m,2}}+2\underbrace{\frac{\text{Re}({\bf h}^*\boldsymbol{z})}{N}\sqrt{p_m}}_{w_{m,3}}\nonumber\\
	&=\underbrace{p_m+\sigma_{z}^2}_{\mu_{m}} + \underbrace{w_{m,1}+w_{m,2}+w_{m,3}}_{w_m}\label{received_energy_model}\\
	&=\mu_{m}+w_m,\nonumber
\end{align}
\endgroup
where $w_m$ represents the deviation of $\widetilde{y}$ from $\mu_{m}$ due to empirical averages. Therefore, we consider $w_m$ as the noise term in our system. Now, invoking the central limit theorem and the law of large numbers,\footnote{Let $\left\{x_1,\cdots,x_n\right\}$ be a sequence of independent and identically distributed random variables with mean $\mu_x$ and variance $\sigma_x^2$. Defining $X=\sum_{i=1}^{n}x_{n}$ and employing central limit theorem along with the law of large numbers, we have $\lim_{n\to\infty}\frac{X-n\mu_{x}}{\sqrt{n}}\sim\mathcal{N}(0,\sigma_x^2)$\cite{rosenthal2006first}.} we show that when $N$ is very large, $w_m$ becomes a zero-mean Gaussian random variable with variance $\sigma_{m}^{2}=\frac{2K+1}{N(K+1)^2}p_m^2+\frac{\sigma_{z}^4}{N}+\frac{2\sigma_{z}^2}{N}p_m$, i.e., $w_{m}\sim\mathcal{N}(0,\sigma_{m}^{2})$. Hence, we have $\widetilde{y}\sim\mathcal{N}(\mu_{m},\sigma_{m}^{2})$. It is clear that $\sigma_{1}^{2}<\sigma_{2}^{2}<\cdots<\sigma_{M}^{2}$. As seen in (\ref{received_energy_model}), when we have a sufficiently large number of antennas at the receiver, we can characterize the average received energy with the transmitted symbol power $p_{m}$, the channel parameters $K$ and $\sigma_{z}^2$, and the number of antennas $N$. Hence, assuming that the receiver applies a MAP detector, we have
\begin{equation}\label{detection}
\widehat{x}=x_{k}\text{ where }k=\arg\!\max_{m\in\{1,\cdots,M\}}f\{\widetilde{y}|x_{m}\},
\end{equation}
where $\widehat{x}$ is the detector output. Given that $x_{m}$ is transmitted, the conditional probability density function (pdf) of $\widetilde{y}$ is
\begin{equation}\label{pdf}
f\{\widetilde{y}|x_{m}\}=\frac{1}{\sqrt{2\pi\sigma_{m}^{2}}}\exp\left(-\frac{(\widetilde{y}-\mu_{m})^{2}}{2\sigma_{m}^{2}}\right).
\end{equation}
Noting that $\widetilde{y}$ is a positive real number, the receiver divides the positive real line into $M$ non-overlapping decision regions: $D_{1},\cdots,D_{M}$, where $D_{m}$ corresponds to the decision region of $x_{m}$. Following (\ref{detection}) and (\ref{pdf}), we can easily infer that the decision regions are defined as $D_{m}=[\lambda_{m-1},\lambda_{m}]$ for $\exists$ $\lambda_{m}\in[\mu_{m},\mu_{m+1}]$ as depicted in Fig. \ref{Received_Constellation}. By default, we set $\lambda_{0}=0$ and $\lambda_{M}=\infty$. Now, we can easily determine the symbol error probability.
Initially, let us focus on the symbols that are at the ends of the constellation diagram. Assume that $x_{1}$ is transmitted. The receiver will be able to decode it correctly when $\widetilde{y}\leq\lambda_{1}$. Otherwise, the receiver output will be wrong. Having the conditional pdf, we can express the symbol error probability when $x_{1}$ is transmitted as $P_{e,1}=\int_{\lambda_{1}}^{\infty}f\{\widetilde{y}|x_{1}\}d\widetilde{y}=Q\left(\frac{\lambda_{1}-\mu_{1}}{\sqrt{\sigma_{1}^{2}}}\right)$, where $Q(x)=\frac{1}{\sqrt{2\pi}}\int_{x}^{\infty}e^{-\frac{t^{2}}{2}}dt$ is the Q-function. On the other hand, when $x_{M}$ is transmitted, the receiver will decode correctly only when $\lambda_{M-1}\leq\widetilde{y}$. Then, the symbol error probability is given by
\begin{align*}
P_{e,M}&=1-\int_{\lambda_{M-1}}^{\infty}f\{\widetilde{y}|x_{M}\}d\widetilde{y}\\
&=1-Q\left(\frac{\lambda_{M-1}-\mu_{M}}{\sqrt{\sigma_{M}^{2}}}\right)=Q\left(\frac{\mu_{M}-\lambda_{M-1}}{\sqrt{\sigma_{M}^{2}}}\right).
\end{align*}
As for the symbols that are located in between $x_{1}$ and $x_{M}$, we can easily see that when $x_{m}$ is transmitted for $m\in\{2,\cdots,M-1\}$, we have the symbol error probability
\begingroup
\allowdisplaybreaks
\begin{align*}
P_{e,m}&=1-\int_{\lambda_{m-1}}^{\lambda_{m}}f\{\widetilde{y}|x_{m}\}d\widetilde{y}\\
&=Q\left(\frac{\mu_{m}-\lambda_{m-1}}{\sqrt{\sigma_{m}^{2}}}\right)+Q\left(\frac{\lambda_{m}-\mu_{m}}{\sqrt{\sigma_{m}^{2}}}\right).
\end{align*}
\endgroup
Since each symbol is transmitted with probability $\frac{1}{M}$, we can express the average symbol error probability as
\begin{align}\label{sym_err_pro}
\hspace{-0.1cm}P_{e}=&\frac{1}{M}\sum_{m=1}^{M}P_{e,m}\nonumber\\
=&\frac{1}{M}\sum_{m=1}^{M-1}\left[Q\left(\frac{\lambda_{m}-\mu_{m}}{\sqrt{\sigma_{m}^{2}}}\right)+Q\left(\frac{\mu_{m+1}-\lambda_{m}}{\sqrt{\sigma_{m+1}^{2}}}\right)\right].
\end{align}
Given a power allocation vector $\mathbf{p} = [p_{1},\cdots,p_{M}]$, we can immediately notice that the objective function in (\ref{sym_err_pro}), $P_{e}$ is separable into $M-1$ sub-functions, i.e.,
\begin{equation*}
P_{e}=\frac{1}{M}\sum_{m=1}^{M-1}h(\lambda_{m})
\end{equation*}
where
\begin{equation}\label{func_mu}
h(\lambda_{m})=Q\left(\frac{\lambda_{m}-\mu_{m}}{\sqrt{\sigma_{m}^{2}}}\right)+Q\left(\frac{\mu_{m+1}-\lambda_{m}}{\sqrt{\sigma_{m+1}^{2}}}\right).
\end{equation}
where $\mu_{m}\leq\lambda_{m}\leq\mu_{m+1}$. We can show that the function $h(\lambda_{m})$ in (\ref{func_mu}) is convex with respect to $\lambda_{m}$ in its defined range $\left[\mu_{m},\mu_{m+1}\right]$. Hence, by taking the derivative of $h(\lambda_{m})$ with respect to $\lambda_{m}$ and equating the derivative to zero, we can easily obtain the optimal $\lambda_{m}$ that minimizes $h(\lambda_{m})$. Hence, the optimal boundary defined with $\lambda_{m}$ between two consecutive decision regions (i.e., the regions of $p_{m}$ and $p_{m+1}$) is given by
\begin{align}\label{lambda_ve}
\lambda_{m}&=\sigma_{z}^{2}+\frac{p_{m}\sigma_{m+1}^{2}-p_{m+1}\sigma_{m}^{2}}{\sigma_{m+1}^{2}-\sigma_{m}^{2}}\\
&\hspace{-0.8cm}+\frac{\sqrt{\sigma_{m}^{2}\sigma_{m+1}^{2}\left\{\left(p_{m+1}-p_{m}\right)^{2}+\left(\sigma_{m+1}^{2}-\sigma_{m}^{2}\right)\log\frac{\sigma_{m+1}^{2}}{\sigma_{m}^{2}}\right\}}}{\sigma_{m+1}^{2}-\sigma_{m}^{2}}.\nonumber
\end{align}

\section{Optimal Power Allocation}
In this section, we provide the optimal constellation diagram that minimizes the average symbol error probability (\ref{sym_err_pro}) under the average power constraint (\ref{int_constraint}) for a given constellation size, $M$. Now, we have the optimization problem as
\begin{equation}
\label{min_prob}
\mathbf{p}^{\star}=\arg\!\min_{\mathbf{p}}P_{e}, \quad \text{s.t.} \quad \frac{1}{M}\sum_{m=1}^{M}p_{m}=\bar{p},
\end{equation}
where $\mathbf{p}^{\star}$ is the vector that holds the power levels of the carriers of the optimal constellation diagram. The above optimization problem does not hold a closed-form solution, and it is in general an NP-hard problem. Therefore, we resort to an iterative algorithm that converges to the optimal solution. Furthermore, regarding the convexity of $P_{e}$ as a function of $\mathbf{p}$, we provide the following result: 

\begin{prop}\label{theo:convexity}
The average symbol error probability $P_{e}$ given in (\ref{sym_err_pro}) is convex in the space spanned by $\mathbf{p}$.
\end{prop}

\emph{Proof:} Omitted due to the space constraints. $\hfill{\square}$

In the sequel, we develop a stepwise algorithm to handle the minimization problem in (\ref{min_prob}). Initially, let us express $P_{e}$ as a function of $\mathbf{p} = [p_{1},\cdots,p_{M}]$ and $\boldsymbol{\lambda}=[\lambda_{1},\cdots,\lambda_{M-1}]$ as
\begin{equation}\label{P_e_p_l}
P_{e}(\mathbf{p}, \boldsymbol{\lambda})=\frac{1}{M}\sum_{m=1}^{M}g(p_{m}),
\end{equation}
where
\begin{equation}\label{g_mm}
g(p_{m})=Q\left(\frac{\mu_{m}-\lambda_{m-1}}{\sqrt{\sigma_{m}^{2}}}\right)+Q\left(\frac{\lambda_{m}-\mu_{m}}{\sqrt{\sigma_{m}^{2}}}\right)
\end{equation}
for $m\in\{2,\cdots,M-1\}$,
\begin{equation*}
g(p_{1})=Q\left(\frac{\lambda_{1}-\mu_{1}}{\sqrt{\sigma_{1}^{2}}}\right)\text{ and }g(p_{M})=Q\left(\frac{\mu_{M}-\lambda_{M-1}}{\sqrt{\sigma_{M}^{2}}}\right).
\end{equation*}
Recall that $\mu_{m}$ is defined in (\ref{received_energy_model}). Now, let us assume that we are initially given a power allocation vector ${\bf p}^{(0)}$. Then, we can easily obtain the optimal decision region boundaries in the first step by using (\ref{lambda_ve}), which we denote by $\boldsymbol{\lambda}^{(0)}=[\lambda_{1}^{(0)},\cdots,\lambda_{M-1}^{(0)}]$. Secondly, let us consider that we are given a vector of decision region boundaries $\boldsymbol{\lambda}^{(0)}$. For any given $\boldsymbol{\lambda}$, the optimal solution for (\ref{P_e_p_l}) can be obtained by the Lagrangian method. However, obtaining a closed-form solution for $p_{m}$ is a very difficult task, and not available without numerical analysis. Hence, we follow a different simplified approach, and treat each sub-function $g(p_{m})$ separately. For given $\lambda_{1}$, $g(p_{1})$ will be minimized when $p_{1}$ is set to 0, i.e., $p_{1}=0$. As for the function $g(p_{m})$ defined in (\ref{g_mm}), we can see that $g(p_{m})$ is convex with respect to $p_{m}$ in the defined region $[\lambda_{m-1},\lambda_{m}]$. Then, $g(p_{m})$ will be minimized when
\begin{figure}
    \centering
\includegraphics[width=\figsizee\textwidth]{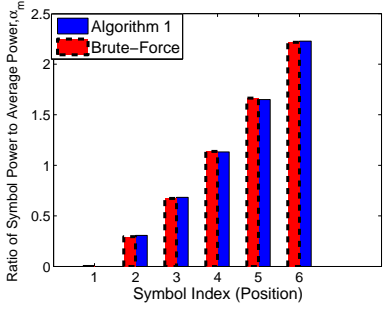}
		\caption{The optimum power allocation diagrams in a given constellation, which are obtained by employing Algorithm 1 and brute-force when $\tSNR = 0$ dB, $N = 500$ and $M = 6$.}\label{fig:fig_2}
\end{figure}
\begin{equation}\label{p_m}
p_{m}=\frac{\lambda_{m}+\lambda_{m-1}}{2}-\sigma_{z}^{2}.
\end{equation}
Subsequently, we can obtain $p_{M}$ by
\begin{equation}\label{pM}
p_{M}=M\bar{p}-\sum_{m=1}^{M-1}p_{m}.
\end{equation}
Then, using $p_{1}=0$, $p_{m}$ in (\ref{p_m}) and $p_{M}$ in (\ref{pM}), we form ${\bf p}^{(1)}$ in the second step. In the case of $p_{M}$ taking any value between $p_{m-1}$ and $p_{m}$, i.e., $p_{m-1}<p_{M}<p_{m}$ for $m\in\{2,\cdots,M-1\}$, we reorganize ${\bf p}^{(1)}$ as ${\bf p}^{(1)}=[p_{1},\cdots,p_{m-1},p_{M},p_{m},\cdots,p_{M-1}]$. Then, we continue the algorithm until we reach a solution that satisfies the termination conditions. In order to formulate, we present our algorithm as
\begin{align*}
{\bf p}^{(0)}\to\boldsymbol{\lambda}^{(0)}\to{\bf p}^{(1)}\to\boldsymbol{\lambda}^{(1)}\to\cdots\to{\bf p}^{(\star)}.
\end{align*}
In the following, we wrap up the above solution into an iterative algorithm:
\begingroup
\captionof{algorithm}{Optimal power allocation\label{algo:power}}
\begin{algorithmic}[1]
\State Set small $\epsilon$ as a stopping criterion where $0<\epsilon$;
\State Initialize $\mathbf{p}=[p_{1},\cdots,p_{M}]$ and set $p_{1}=0$ such that the average power constraint in (\ref{int_constraint}) is satisfied, and $p_{1}<p_{2}<\cdots<p_{M}$;
\While{True}
\State Given $\mathbf{p}$, compute $\boldsymbol{\lambda}=[\lambda_{0},\cdots,\lambda_{M}]$ using (\ref{lambda_ve});
\State Given $\boldsymbol{\lambda}$, compute $\mathbf{p}^{\star}=[p_{1}^{\star},\cdots,p_{M}^{\star}]$ by using (\ref{p_m}) and (\ref{pM});
\If{$p_{m-1}^{\star}\leq p_{M}^{\star}\leq p_{m}^{\star}$ for any $m$}
\State Set $\mathbf{p}^{\star}=[p_{1}^{\star},\cdots,p_{m-1}^{\star},p_{M}^{\star},p_{m}^{\star},\cdots,p_{M-1}^{\star}]$
\EndIf
\If{$\|\mathbf{p}-\mathbf{p}^{\star}\|^{2}<\epsilon$}
\State \textbf{break};
\Else
\State Set $\mathbf{p}=\mathbf{p}^{\star}$;
\EndIf
\EndWhile
\end{algorithmic}
\endgroup
\begin{figure}
    \centering
    \subfigure[K=50]
    {\includegraphics[width=\figsizee\textwidth]{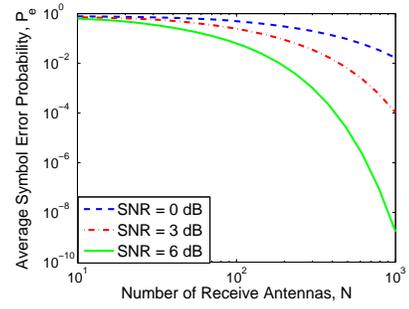}\label{fig:fig_3_a}}
    \subfigure[K=0]
    {\includegraphics[width=\figsizee\textwidth]{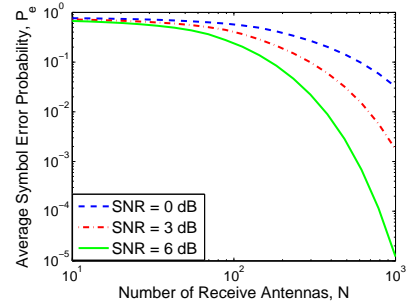}\label{fig:fig_3_b}}
		\caption{The average symbol error probability, $P_{e}$, v.s. the number of receive antennas, $N$, for $M = 10$ and different $\tSNR$ and $K$ values.}\label{fig:fig_3}
\end{figure}

The above algorithm reaches the optimal solution after 6 iterations when the constellation size is $M = 4$ and the number of receive antennas is $N = 500$, and it requires about 15 iterations to reach the solution when $M = 6$ with the same number of receive antennas.
\begin{figure*}[ht!]
    \centering
    \subfigure[K=50]
    {\includegraphics[width=\figsizee\textwidth]{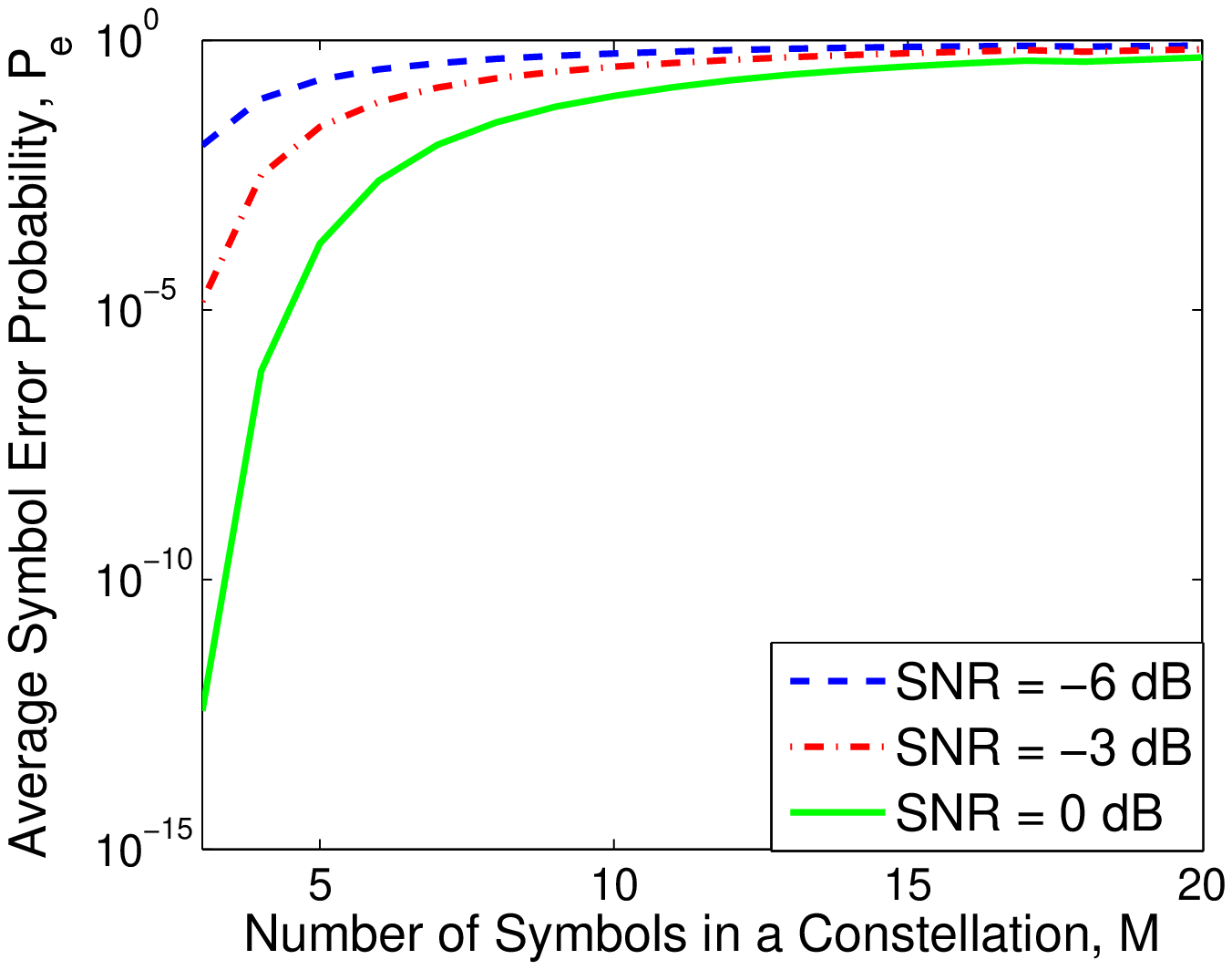}\label{fig:fig_4_a}}
    \subfigure[K=0]
    {\includegraphics[width=\figsizee\textwidth]{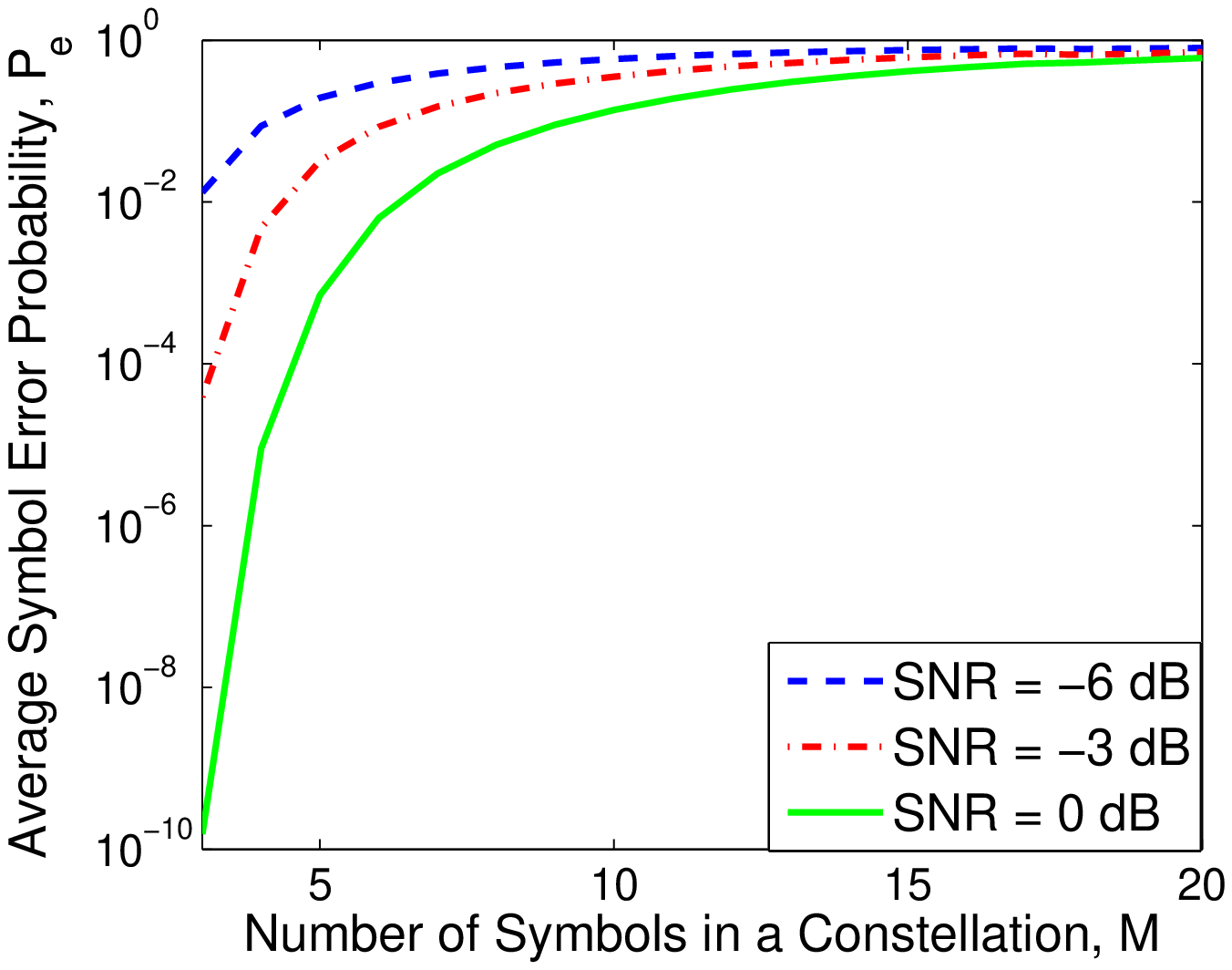}\label{fig:fig_4_b}}
		\caption{The average symbol error probability, $P_{e}$, v.s. the number of symbols in a given constellation, $M$, for $N = 500$ and different $\tSNR$ and $K$ values.}\label{fig:fig_4}
\end{figure*}

\begin{figure*}[ht!]
    \centering
    \subfigure[K=50]
    {\includegraphics[width=\figsizee\textwidth]{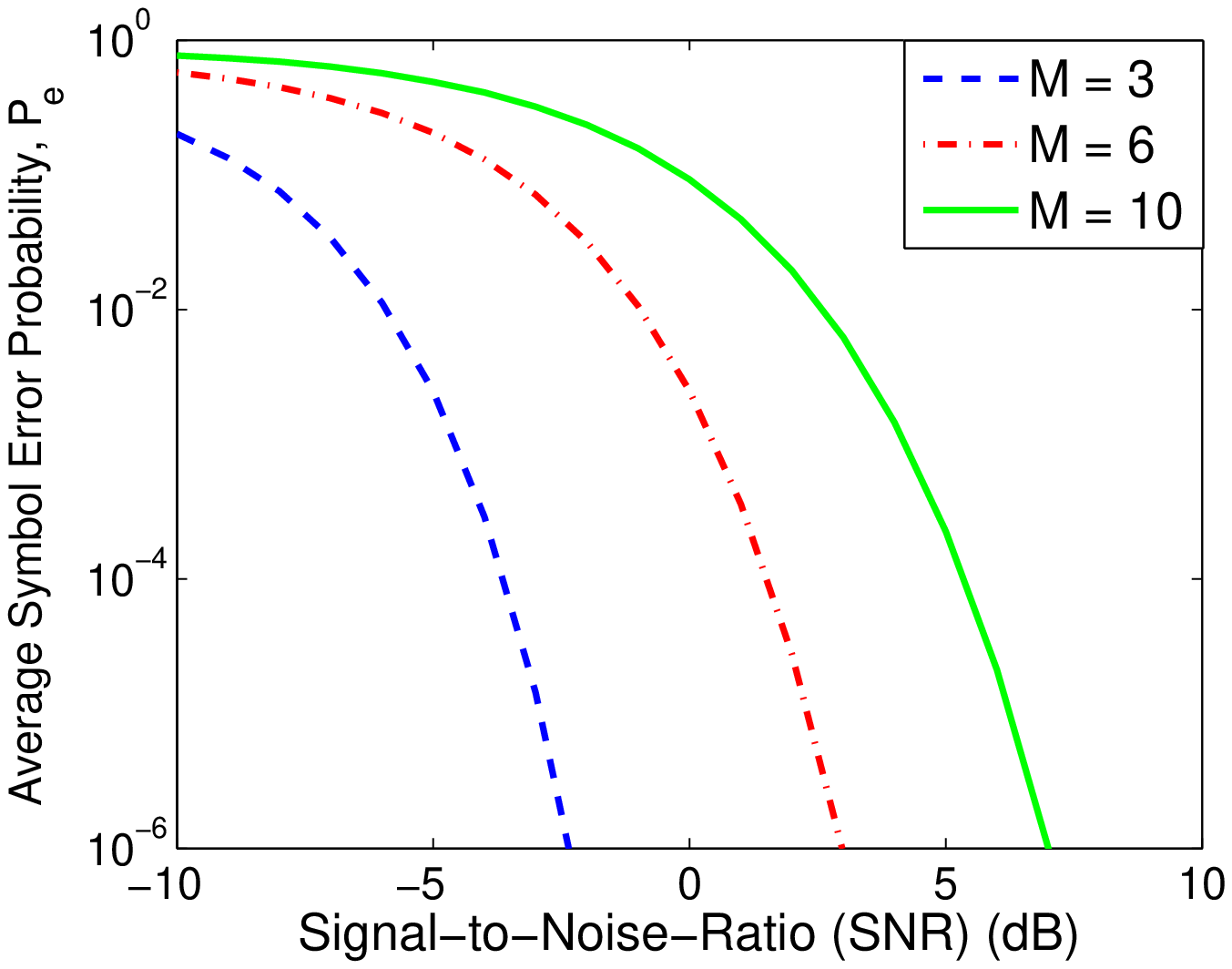}\label{fig:fig_8_a}}
    \subfigure[K=0]
    {\includegraphics[width=\figsizee\textwidth]{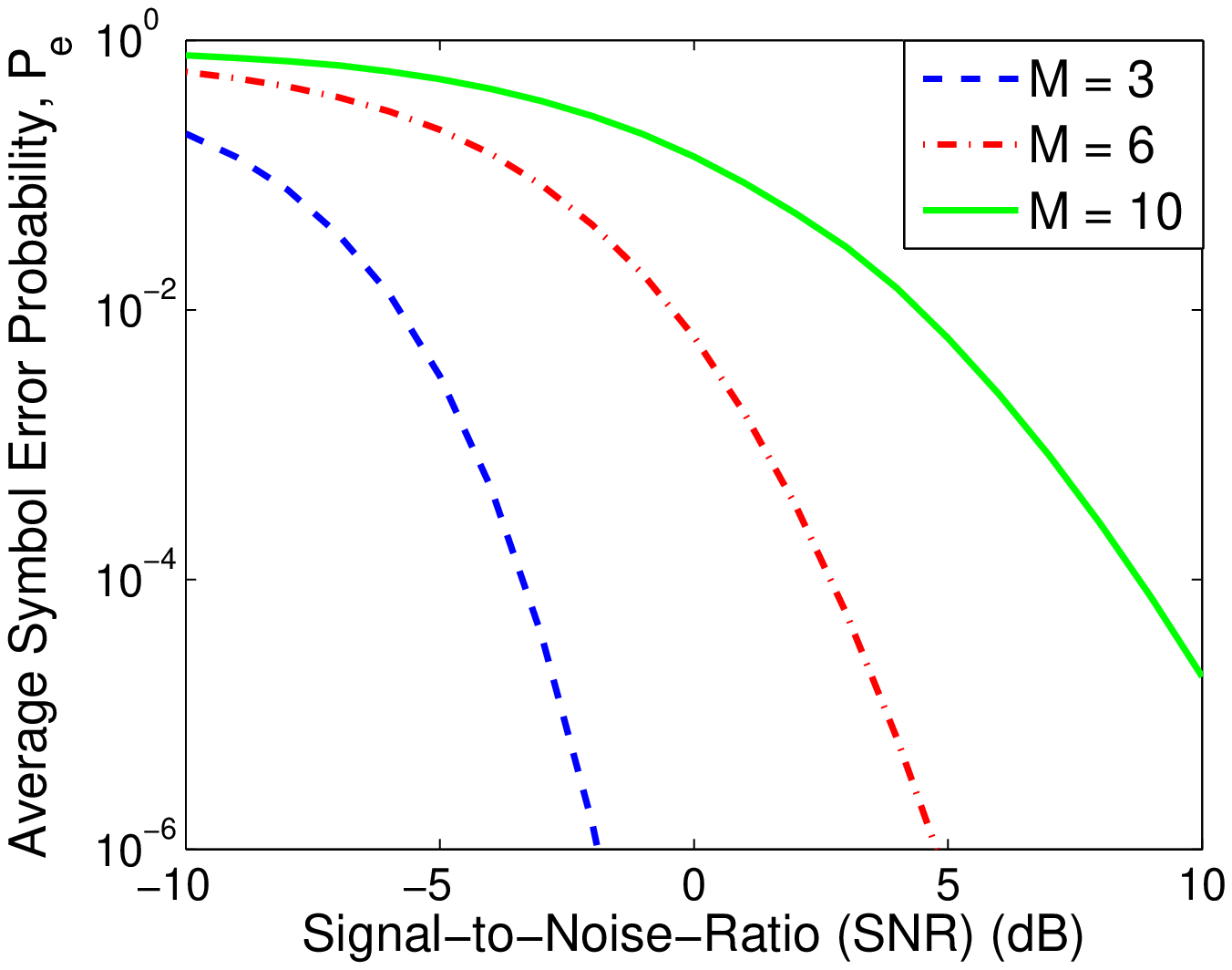}\label{fig:fig_8_b}}
		\caption{The average symbol error probability, $P_{e}$, v.s. the signal-to-noise-ratio (SNR) for $N = 500$ and different $M$ and $K$ values.}\label{fig:fig_8}
\end{figure*}
\section{Numerical Results}
In this section, we present the numerical results. Throughout the paper, we consider the following settings and parameters unless specified otherwise. In each figure, except Fig. \ref{fig:fig_2}, we have plots as a pair in one of which we display the results when $K=50$, and in the other the results are obtained regarding a channel when $K=0$. We note that the channel when $K=0$ is considered to be the Rayleigh channel in which there is no dominant propagation along the line-of-sight, while the channel when $K=50$ has a strong line-of-sight propagation. Additionally, we set the stopping criterion $\epsilon=10^{-6}$.

Setting $K=50$ in Fig. \ref{fig:fig_2}, we display the optimal power allocation ratios among the carriers of an optimal constellation diagram using Algorithm 1 when $\tSNR=0$ dB. We further compare the results with those obtained using the optimal Brute-force search. We can easily see the good match between the results. Furthermore, we plot the average symbol error probability as a function of the number of receive antennas, $N$, in Fig. \ref{fig:fig_3} when $\tSNR=0$, $3$, and $6$ dB. Regardless of the line-of-sight character, there is dramatic decrease in the average probability of error with increasing $N$. As expected, the decrease is higher when $K=50$ than it is when $K=0$. Similarly, we plot the average symbol error probability as a function of the number of constellation symbols, $M$, in Fig. \ref{fig:fig_4} for $\tSNR=-6$, $-3$, and $0$ dB in order to see the effects of $M$ when $\tSNR$ is small. We can easily infer that the number of receive antennas has a great impact in obtaining small values of $P_{e}$ even when $\tSNR$ is small especially at lower constellation sizes. Finally, we plot the average symbol error probability, $P_{e}$, as a function of $\tSNR$ in Fig. \ref{fig:fig_8} for different number of symbols in a given constellation, $M$. We see that with increasing $M$, $P_{e}$ increases. We further note that due to a dominant line-of-sight effect we have better results in Fig. \ref{fig:fig_8_a}. 

\section{Conclusion}\label{conclusion}
In this paper, we have investigated the optimal power allocation design for noncoherent energy-detection-based systems in which receivers are furnished with notably many antennas. Under an average transmit power constraint, we have attained the average symbol error probability of a MAP detector, and we have identified the optimal decision regions for this setting. Showing that the average symbol error probability is convex in the space spanned by the symbols of a given constellation diagram, we have provided an iterative algorithm that converges to the optimal power allocation policy among the symbols of this constellation. Through numerical results, we have analyzed the effects of the channel parameters such as the line-of-sight character, the number of receive antennas, and the constellation size on the performance levels.
%

\bibliographystyle{IEEEtran}
\bibliography{references}
\end{document}